\documentclass{elsart1p}
\pdfoutput=1

\usepackage{graphics}
\usepackage{graphicx}
\usepackage{epsfig}

\usepackage{amssymb}

\begin{document}

\begin{frontmatter}

\title{Overview and Perspectives in Nuclear Physics}
\author{Wolfram Weise}\footnote{Supported in part by BMBF, GSI and by the DFG cluster of excellence Origin and Sructure of the Universe.}
\address{Physik-Department, Technische Universit\"at M\"unchen, D-85747 Garching, Germany}



\begin{abstract}
This presentation reviews recent guiding themes\footnote{Of course, such an overview is unavoidably influenced by personal selection. The author apologizes in advance for omissions that are entirely his own responsability.} in the broad context of nuclear physics, from developments in chiral effective field theory applied to nuclear systems, via the phases and structures of QCD, to matter under extreme conditions in heavy-ion collisions and neutron stars.
\end{abstract}

\end{frontmatter}

\section{Introduction}
\label{Intro}
Nuclear physics has broadened its scope and frame steadily over the past few decades \cite{Ri05}. Contemporary nuclear physics is perhaps best characterised as the field of research dealing with the phases and structures of QCD at their various levels of complexity. 

\begin{figure}[htb]
\begin{minipage}[t]{65mm}
\includegraphics[width=6.4cm]{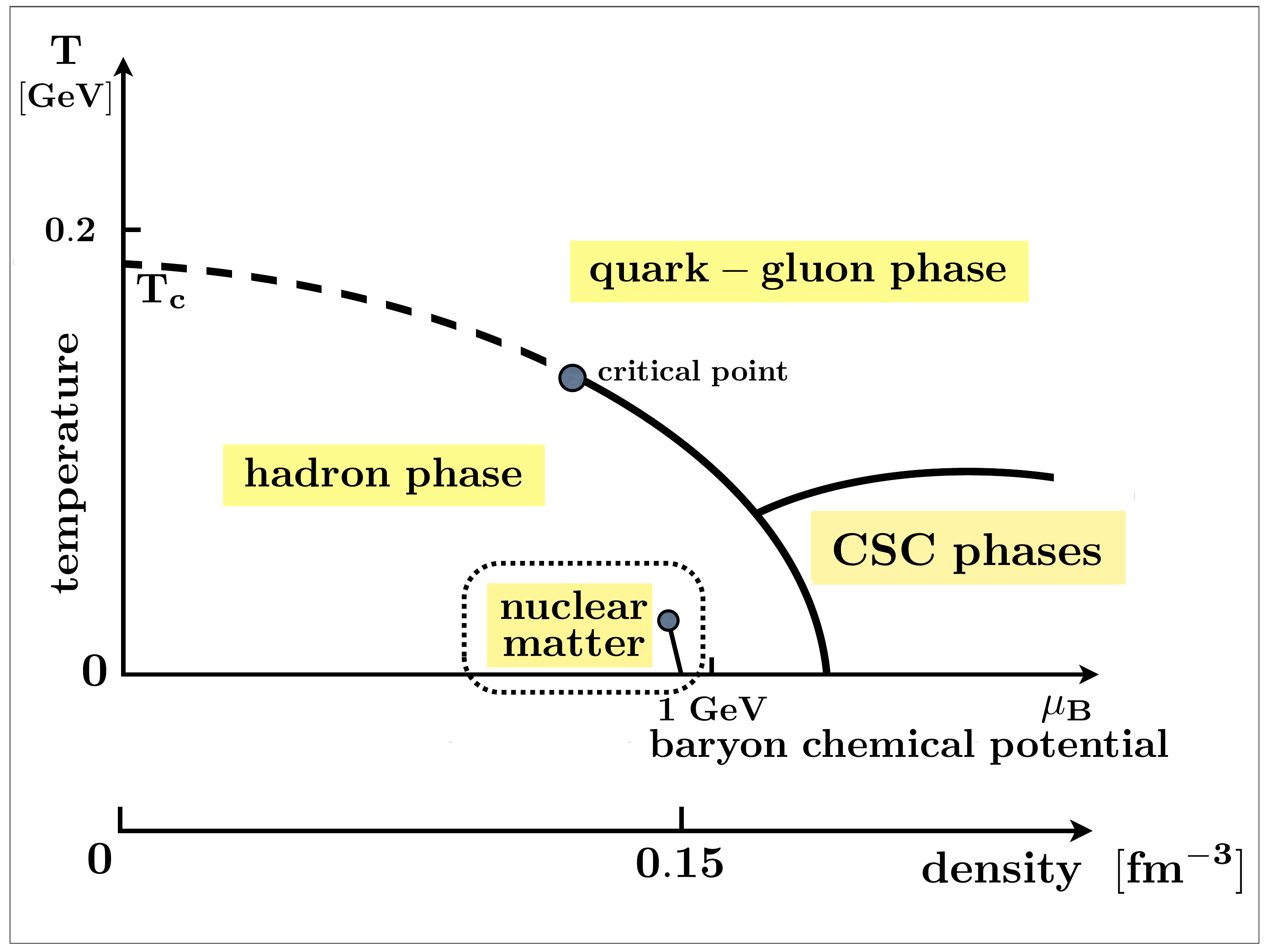}
\caption{Schematic phase diagram of QCD showing the hadronic, quark-gluon and color superconducting (CSC) phases in the plane of temperature and baryon chemical potential. The dashed line marks a crossover, separated from a first order transition by a critical point.}
\label{fig:1}
\end{minipage}
\hspace{\fill}
\begin{minipage}[t]{65mm}
\includegraphics[width=6.4cm]{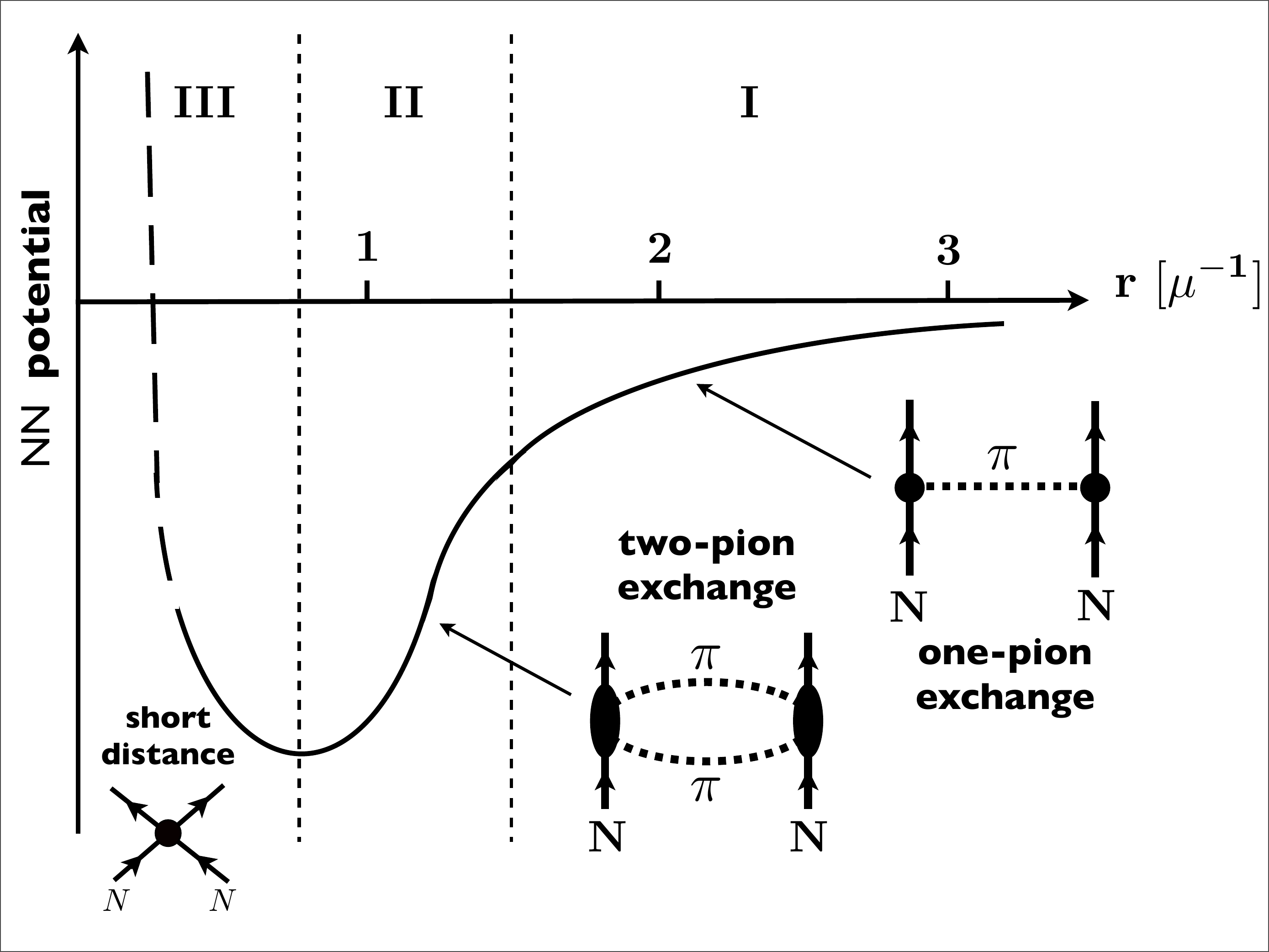}
\caption{Hierarchy of scales governing the nucleon-nucleon interaction (adapted from Taketani  \cite{Tak56}). The distance $r$ is given in units of the pion Compton wavelength, $\mu^{-1} \simeq 1.4$ fm.}
\label{fig:2}
\end{minipage}
\end{figure}

Fig.\ref{fig:1} gives a schematic impression of our expectations concerning the QCD phase diagram. 
In the hadronic phase at low temperatures and densities, quarks and gluons are confined in color singlet composites. Their dynamics is largely governed by the spontaneously broken chiral symmetry of QCD in its sector with almost massless {\it up} and {\it down} quarks.  Nuclei are placed right in the center of the hadronic phase, as aggregates of nucleons and meson fields. Nucleons themselves are complex systems of quarks and gluons. When strongly interacting hadronic matter is heated beyond a critical temperature $T_c \sim 0.2$ GeV, one expects the compounds to dissolve and the hadron phase turns into a phase of liberated quarks and gluons. Precisely how this transition proceeds, and how it relates to the symmetry breaking pattern in QCD, is one of the fundamental questions of modern 
nuclear physics. 

A different class of phenomena is expected to occur at low temperatures but large baryon chemical potentials. At very high quark Fermi momenta, QCD predicts the formation of color-antitriplet diquarks as Cooper pairs. Matter can then presumably exist in color superconducting (CSC) phases. Understanding the mechanisms  which govern the transition from ``normal" nuclear or neutron matter to such extreme baryonic densities is a further fundamental challenge. 

\section{Yukawa's pion, chiral symmetry and low-energy QCD}

This year's Yukawa centennial has inspired retrospectives into the historical origins of nuclear physics. Yukawa's pioneering work of 1935 \cite{Yuk35} laid the foundations for our understanding of the strong force between nucleons. Twenty years later an impressively coherent series of articles appeared: the Collected Papers on Meson Theory \cite{Sup55}. These reviews from half a century ago read, still today, remarkably modern in their conceptual drive towards a systematic approach to the nucleon-nucleon interaction. A cornerstone of these developments was the visionary design by Taketani et al. \cite{TNS51} of the inward-bound hierarchy of scales governing the $NN$ interaction (see Fig.\ref{fig:2}). The long range region I is determined by one-pion exchange. It continues inward to the intermediate distance region II dominated by two-pion exchange. The basic idea is to construct the $NN$ potential in these regions by $explicit$ calculation of $all$ $\pi$ and $2\pi$ exchange processes. The detailed behaviour of the interaction in the short distance region III remains unresolved at the low-energy scales characteristic of nuclear physics. This short distance part is given a suitably parametrized form and fixed by comparison with scattering data. Taketani's early separation-of-scales programme is quite reminiscent of the methods applied today under the heading ``chiral effective field theory"\footnote{For a recent survey from this perspective, see also ref.\cite{WW07}.}. 

QCD can be handled with controlled approximations in two limiting situations:
the limit of high energies and momenta (Q $\gg 1$ GeV) or ultrashort distances (r $<0.1$ fm) in which QCD is a theory of weakly interacting quarks and gluons; and the limit of low energies and momenta (Q $\ll 1$ GeV) or long distances (r $> 1$ fm) in which QCD is realized in the form of an effective field theory of weakly interacting Nambu-Goldstone bosons associated with the spontaneous breaking of chiral symmetry. The vast area between these limits is becoming progressively accessible through lattice QCD, large-scale numerical simulations on discretised Euclidean space-time using the most powerful computers available. 

Low-energy QCD deals with systems of light quarks at energies and momenta smaller than the characteristic scale for spontaneous chiral symmetry breaking, $4\pi f_\pi \sim 1$ GeV (with $f_\pi \simeq 0.09$ GeV the pion decay constant).  The mass gap in the spectrum of the lightest hadrons is a visible manifestation of this scale. It offers a natural separation between {\it light} and {\it heavy} (or, correspondingly, {\it fast} and {\it slow}) degrees of freedom. The basic idea of an effective field theory is to
introduce the light species as active degrees of freedom,  while the
heavy particles are treated as (almost) static sources. In QCD, confinement and spontaneous chiral symmetry breaking imply that the {\it light} degrees of freedom at low energy are color singlet, pseudoscalar Nambu-Goldstone bosons, identified with the pions for $N_f = 2$ quark flavors, and massless in the limit of vanishing quark masses. Their dynamics
is described by an effective Lagrangian which incorporates all relevant
symmetries of the underlying fundamental theory. Explicit chiral symmetry breaking by the small $u$ and $d$ quark masses shifts the pion mass to its physical value, $m_\pi = 0.14$ GeV, one of the {\it small} scales governing nuclear physics. 

The chiral symmetry breaking scenario of low-energy QCD is a basic starting point for today's approaches to nuclear few- and many-body systems. Apart from the identification of Yukawa's pion as an isospin triplet of Nambu-Goldstone bosons, this scenario is characterised by the existence of a strong condensate $\langle\bar{q} q\rangle$ of scalar quark-antiquark pairs (the chiral condensate) in the QCD ground state (the highly non-trivial {\it vacuum}). But is this symmetry breaking scenario unambiguously confirmed? Several recent measurements and observations do indeed give a positive answer to this question. 

Detailed chiral perturbation theory  calculations \cite{CGL01} of low-energy $\pi\pi$ scattering phase shifts have been performed and compared with those extracted from the final state analysis of $K \rightarrow \pi\pi + e\nu$ decays. These results support the ``strong condensate'' scenario of spontaneous chiral symmetry breaking in QCD. It is linked to an important current algebra (PCAC) relation that establishes a leading proportionality between the squared pion mass and the quark mass, $m_\pi^2 \propto m_q$, the slope being determined by the ratio $\langle\bar{q}q\rangle / f_\pi^2$. This relationship is also confirmed by lattice QCD results (unexpectedly even up to quite large quark masses, a feature that is not yet well understood). 

Ongoing tests include the measurement of the electromagnetic polarisabilities of the pion for which there is a definite prediction from chiral perturbation theory \cite{GIS06}. The COMPASS experiment at CERN uses the Primakoff process: scattering of a high energy pion beam (190 GeV) on the electromagnetic field in the periphery of a heavy nuclear target at extreme forward angles, focusing on the smallest possible momentum transfers. The first preliminary results are promising and subject to further detailed analysis. 

\section{Two- and three-nucleon interactions: new developments}

\begin{figure}[htb]
\begin{minipage}[t]{65mm}
\includegraphics[width=6.5cm]{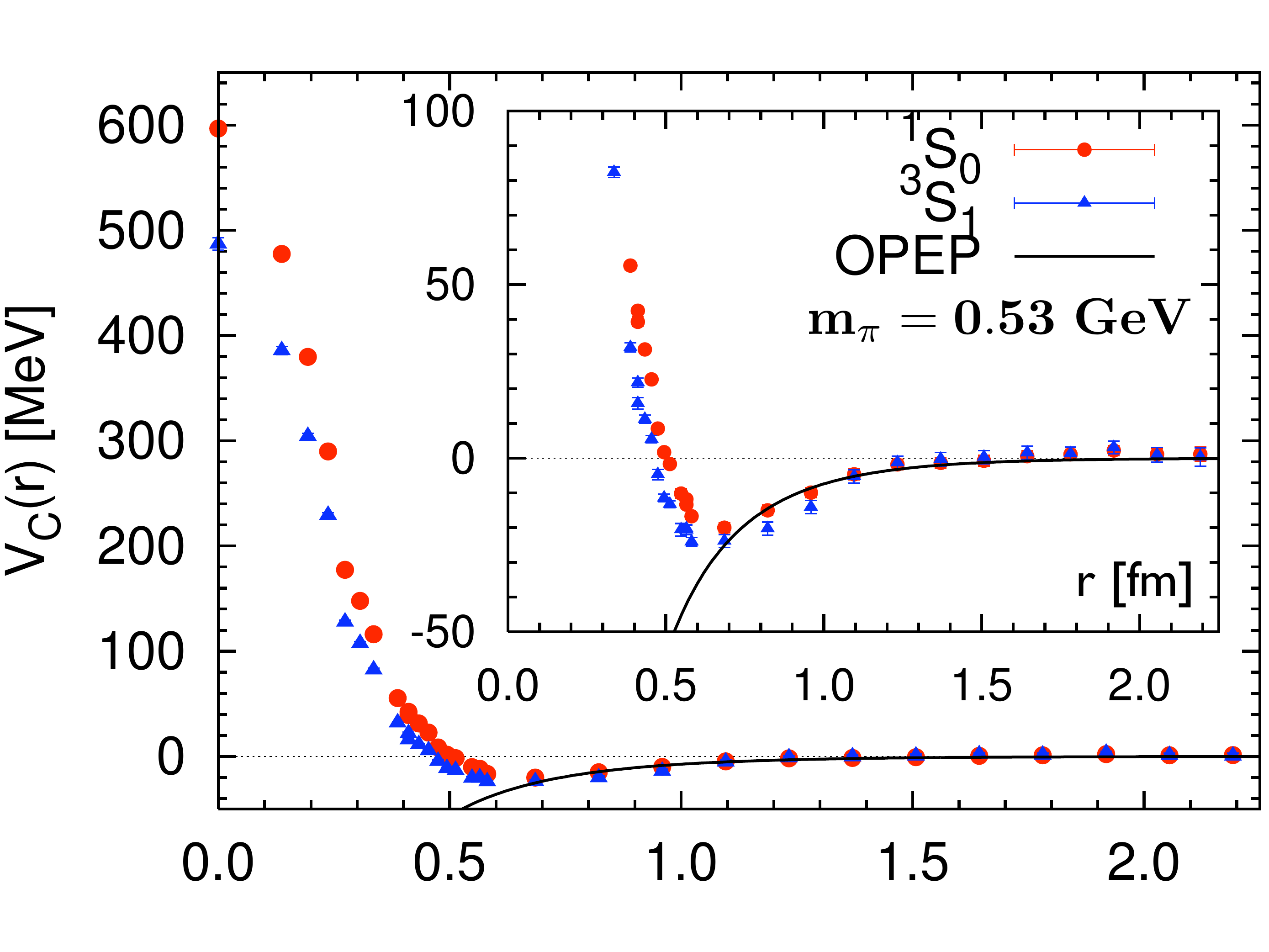}
\caption{Central nucleon-nucleon potential deduced from lattice QCD \cite{IAH07}. The solid curve shows a comparison with one-``pion" exchange with a pion mass adapted to the quark mass used in the lattice computations.}
\label{fig:3}
\end{minipage}
\hspace{\fill}
\begin{minipage}[t]{65mm}
\includegraphics[width=6.4cm]{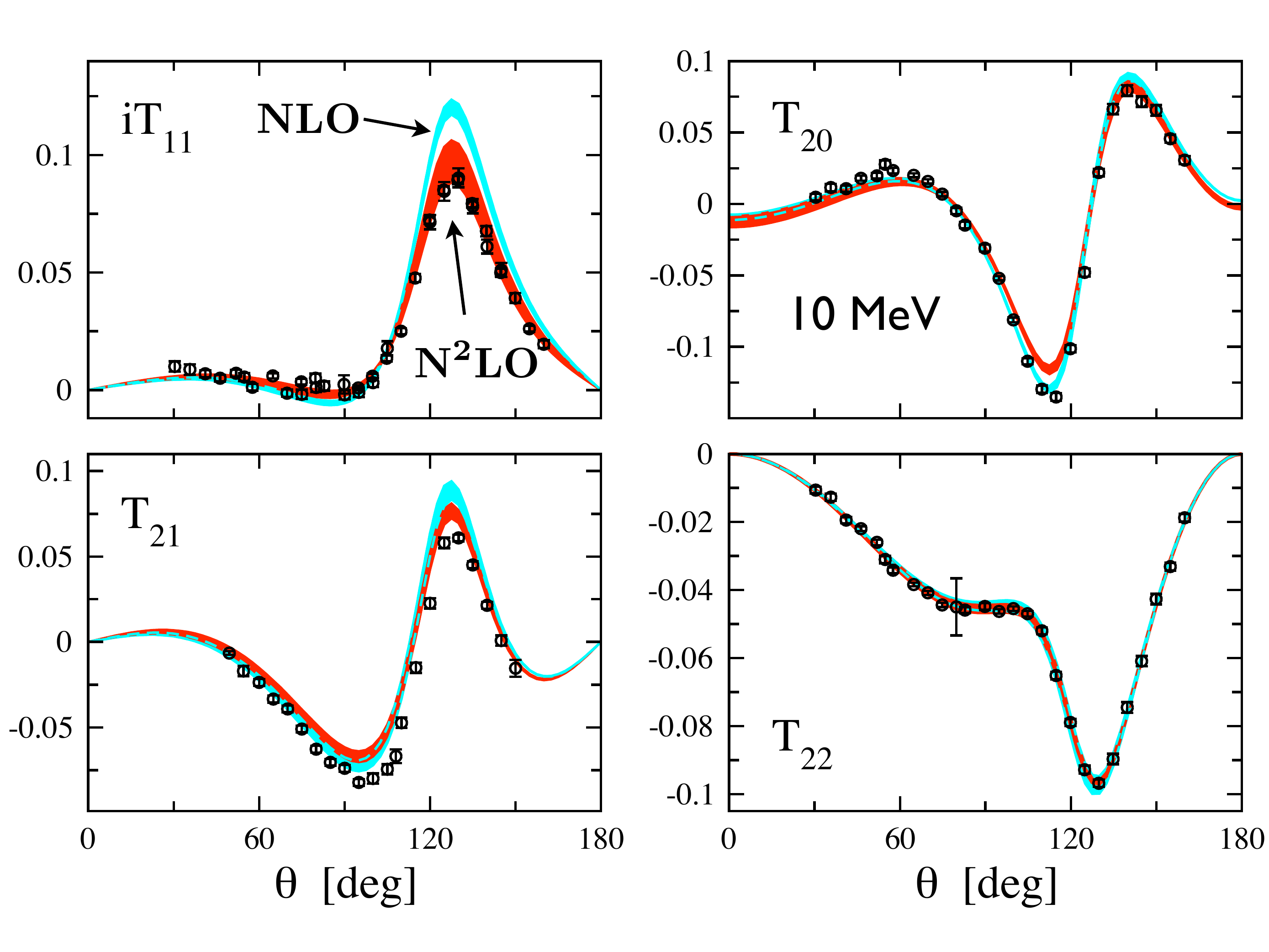}
\caption{Tensor analysing powers for elastic $nd$ scattering at 10 MeV. Next-to-leading order and
next-to-next-to-leading order results of a chiral effectice field theory calculation are shown in comparison with experimental data. Adapted from \cite{Ep06}.}
\label{fig:4}
\end{minipage}
\end{figure}

3.1~{\it NN interaction from lattice QCD}.
Recent lattice studies of the nucleon-nucleon force \cite{IAH07} are beginning to draw a QCD based picture of the $NN$ potential. These computations first generate the wave function $\Phi(r)$ of two nucleons at a distance $r$ on the lattice. The central potential is then reconstructed from the Schr\"odinger equation as $V_c(r) = E + {\nabla^2 \Phi(r)\over 2\mu\,\Phi(r)}$. The resulting $V_c(r)$, shown in Fig.\ref{fig:3}, displays indeed the known qualitative features of the $NN$ interaction: a short-range repulsive core followed by attraction at intermediate and long distances, with a splitting into singlet ($^1 S_0$) and triplet ($^3 S_1$) channels. This is evidently not yet a realistic potential since the result is obtained in quenched QCD and the quark masses used are still too large, equivalent to pion masses several times the physical $m_\pi$. But it is the promising starting point of a development which provides, in particular,  a foundation in QCD for the repulsive short-distance dynamics of the $NN$ force. 

3.2~{\it Nuclear interactions from chiral effective field theory}.
The separation of scales characteristic of chiral effective field theory defines a systematic hierarchy of contributions to the nucleon-nucleon interaction and nuclear three-body forces, driven by pions as Nambu-Goldstone bosons.  The emerging series is organised in powers of the small quantity $Q/4\pi f_\pi$, where $Q$ stands generically for low energy or momentum  and $4\pi f_\pi$ is the spontaneous chiral symmetry breaking scale of order 1 GeV. The familiar one-pion exchange interaction comes at leading order. At next-to-leading order, ${\cal O}(Q^2)$, a first set of two-pion exchange mechanisms enters together with contact terms encoding unresolved short distance dynamics. At the next higher order (${\cal O}(Q^3)$), more two-pion exchange processes are turned on, in particular those involving the strong spin-isospin polarizablity of the nucleon as it is manifest in the $N\rightarrow \Delta(1232)$ transition that dominates $p$-wave pion-nucleon scattering. At that same order three-body interactions have their entry in a well defined book-keeping scheme.

With a few constants fixed by comparison with $NN$ and three-nucleon data, this approach is being applied successfully to few-body systems (see ref.\cite{Ep06} for a recent review).
Fig.\ref{fig:4} shows, as one out of many impressive examples, the calculated tensor analysing powers in low-energy elastic neutron-deuteron scattering confronted with high-precision measurements. 

\section{From QCD via effective field theory to the nuclear chart}

\begin{figure}[htb]
\begin{minipage}[t]{65mm}
\includegraphics[width=6.5cm]{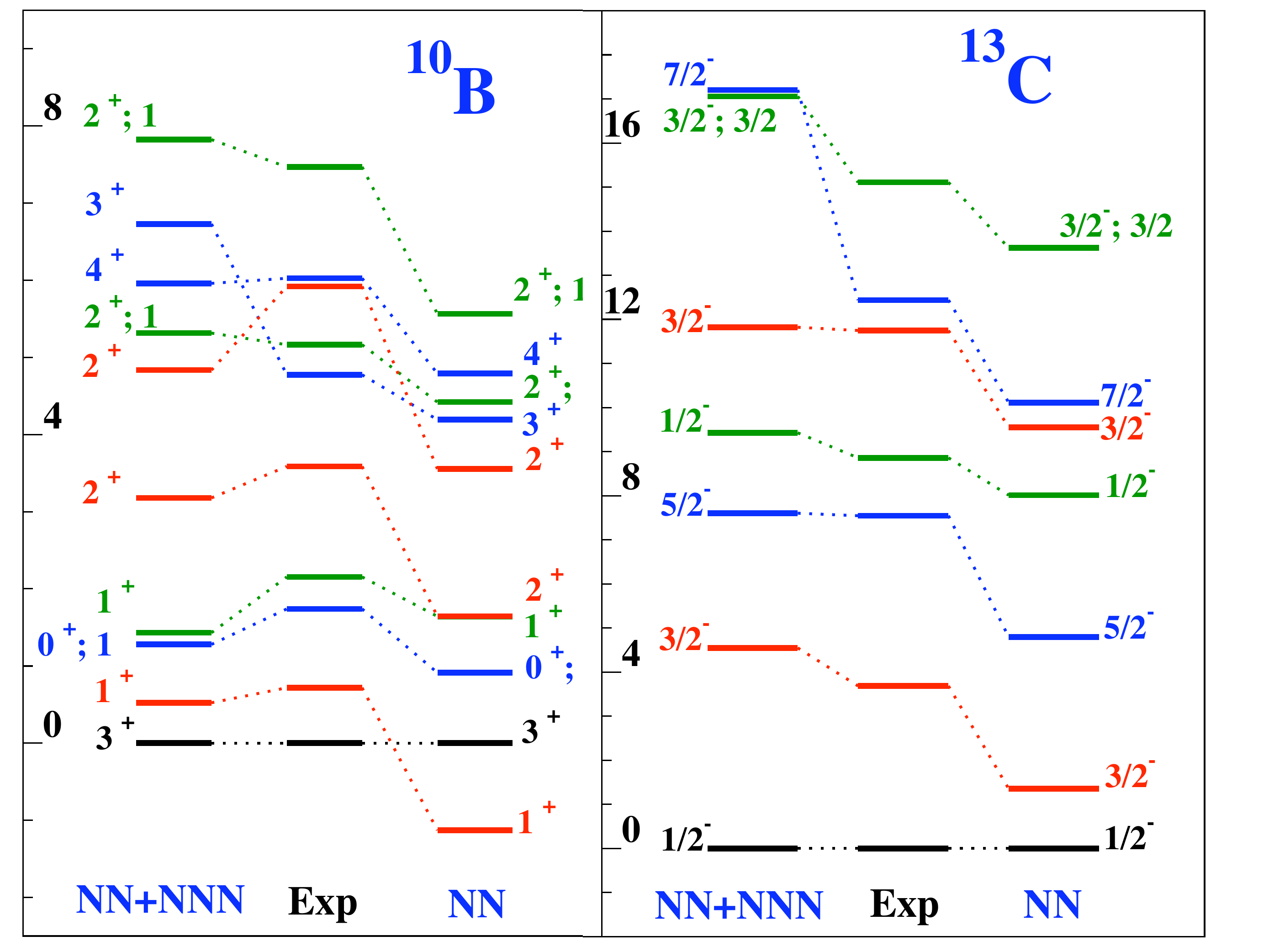}
\caption{Examples of no-core shell model calculations for p-shell nuclei \cite{Na07} using $NN$ and $NNN$ interactions based on chiral effective field theory. The left columns in each subfigure show the effects of three-nucleon forces added to the $NN$ interaction, in comparison with experimental spectra (in MeV).}
\label{fig:5}
\end{minipage}
\hspace{\fill}
\begin{minipage}[t]{65mm}
\includegraphics[width=6.5cm]{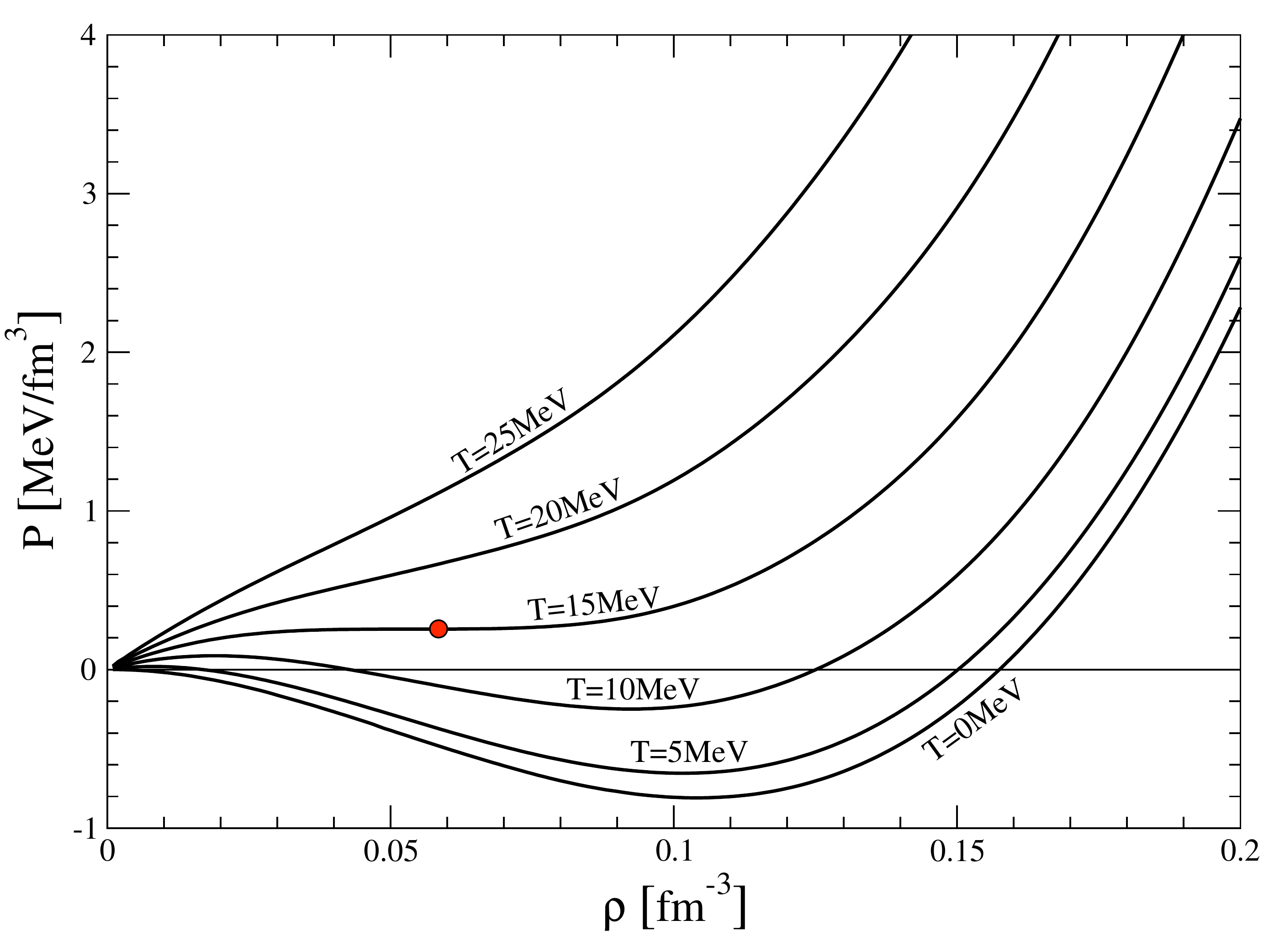}
\caption{Equation of state of nuclear matter: pressure $P$ as function of density $\rho$ for different temperatures $T$. The calculation is performed using in-medium chiral effective field theory \cite{FKW05}. The point indicates a first-order liquid-gas transition at a critical temperature $T_c \simeq 15$ MeV.}
\label{fig:6}
\end{minipage}
\end{figure}

4.1~{\it Example: p-shell nuclei}. The important role of three-body forces driven by tensor interactions from pion exchange and involving prominently the $\Delta(1232)$ isobar  has long been emphasised by the late V.J. Pandharipande and his collaborators \cite{PP01}. Extensive no-core shell model calculations can now be performed using $NN$ and $NNN$ interactions derived from chiral effective field theory, including explicit two-pion exchange dynamics. Representative examples for p-shell nuclei \cite{Na07} are shown in Fig.\ref{fig:5}. They demonstrate the improvement in understanding the spectra of low-lying excited states of A = 10-13 nuclei when incorporating chiral $NNN$ forces consistent with three-nucleon data.

4.2~{\it Chiral dynamics and the nuclear many-body problem}.  In nuclear matter an additional relevant momentum scale is the Fermi momentum $p_F$. Around the empirical saturation point with $p_F \simeq 0.26$ GeV, the nucleon Fermi momentum  and the pion mass, together with the $N-\Delta$ mass difference, are all comparable {\it small} scales: we have $p_F \sim 2\,m_\pi \sim M_\Delta - M_N \ll 4\pi\,f_\pi \sim 1$ GeV. This implies that at the densities of interest in nuclear physics, $\rho \lesssim\rho_0 = 2p_F^3/3\pi^2 \simeq 0.16$ fm$^{-3} \simeq 0.45\,m_\pi^3$, pions $must$ be treated as $explicit$ degrees of freedom in any meaningful description of the nuclear many-body problem. The strong pion-exchange tensor force and, in particular, two-pion exchange processes involving intermediate spin-isospin  ($N\rightarrow \Delta$) excitations, play a leading role at the distance scales characteristic of the nuclear bulk. Although these basic principles were known for many decades, it took an astonishingly long time in nuclear many-body theory until chiral two-pion exchange dynamics was taken seriously  - rather than just being parametrised in terms of ($\sigma$ and $\rho$) boson exchange phenomenology.  
     
In-medium chiral perturbation theory has emerged as a successful framework for low-energy pion-nucleon dynamics in the presence of a filled Fermi sea of nucleons. One- and two-pion exchange processes, treated explicitly, govern the long-range interactions at distance scales $d \gtrsim 1/p_F$ relevant to the nuclear many-body problem. Short-range mechanisms, with spectral functions involving masses far beyond those of two pions, are not resolved in detail at nuclear Fermi momentum scales and can be subsumed in contact interactions and derivatives thereof. This {\it separation of scales} argument makes strategies of chiral effective field theory work also for nuclear problems, with the {\it small} scales ($p_F, m_\pi, M_\Delta - M_N $) distinct from the {\it large} ones ($4\pi f_\pi, M_N$). In essence, this is the modern realisation of Taketani's programme mentioned earlier. Closely related renormalisation group considerations have motivated the construction of a universal low-momentum $NN$ interaction V($low\, k$) \cite{BKS03} from phase shift equivalent $NN$ potentials such that the ambiguities associated with unresolved short-distance parts disappear. 

The prominent pieces of the $2\pi$ exchange interaction involve the second order tensor force and intermediate $\Delta(1232)$ states. The latter produces a Van der Waals - like $NN$ interaction which behaves, at long and intermediate distances, as $V_{2\pi}(r) \propto -e^{-2m_\pi r}/r^6$ times a polynomial in $m_\pi r$. It does then perhaps not come as a surprise that the resulting nuclear matter equation of state \cite{FKW05}, see Fig.\ref{fig:6}, is qualitatively reminiscent of a Van der Waals equation of state. The nuclear liquid turns into a gas at a calculated critical temperature $T_c \simeq 15$ MeV, quite close to the commonly accepted empirical range $T_c \sim 16-18$ MeV.

4.3~{\it Towards heavy nuclei and beyond}.  A successful strategy for approaching finite nuclei over a broad range, from $^{16}O$ to the very heavy ones, starts from a universal (relativistic) energy density functional guided by nuclear matter calculations. The basic idea is to construct the exchange correlation part of this functional from in-medium chiral perturbation theory, with contact (mean field) terms added. In this approach binding and saturation, for nuclear matter as well as finite nuclei, are driven primarily by attractive two-pion exchange mechanisms in combination with repulsive Pauli principle effects. Emprical properties of nuclei (binding energies, charge radii etc.) 
are reproduced throughout the nuclear chart \cite{FKVW06}, typically within 0.5\%. This is a level of precision comparable to the best phenomenological mean field calculations, but now with constraints imposed by known principles of low-energy QCD. 

\begin{figure}[htb]
\begin{minipage}[t]{63mm}
\includegraphics[width=6.3cm]{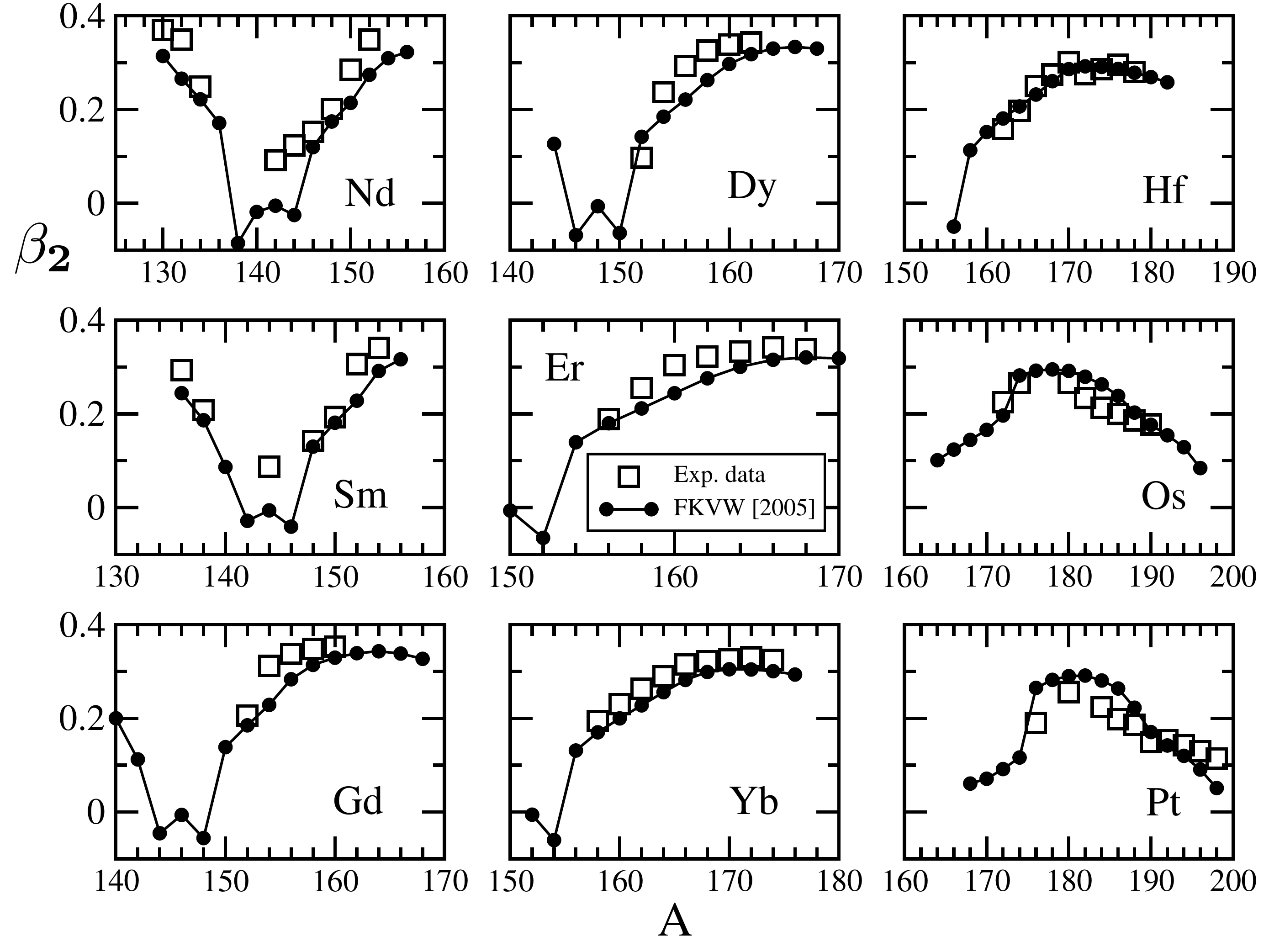}
\caption{Ground state quadrupole deformations of a series of isotopes calculated in a relativistic Hartree-Bogoliubov model using interactions derived from in-medium chiral perturbation theory \cite{FKVW06}, compared with experimental data (squares).}
\label{fig:7}
\end{minipage}
\hspace{\fill}
\begin{minipage}[t]{67mm}
\includegraphics[width=6.7cm]{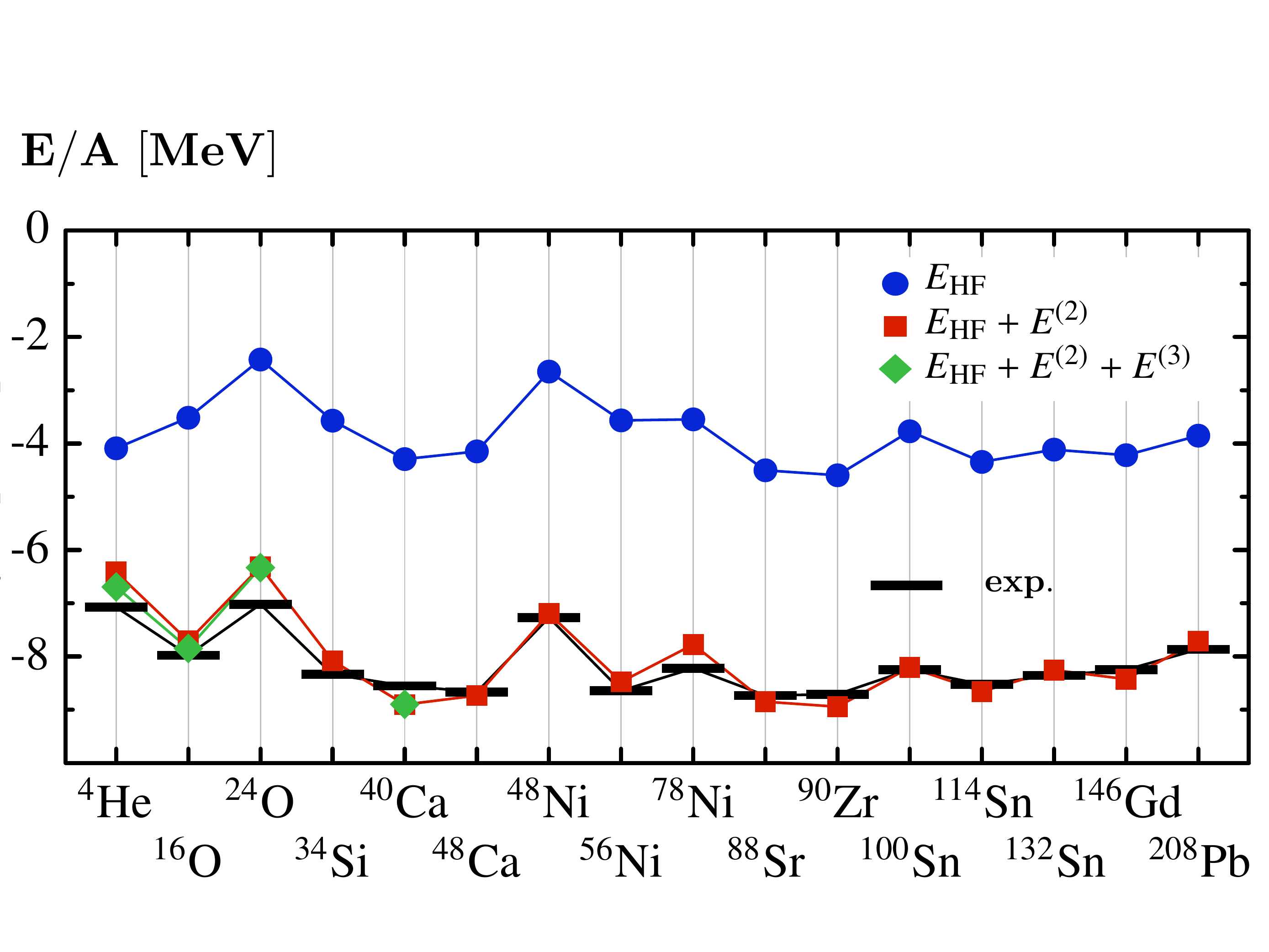}
\caption{Ground state energies of selected closed-shell nuclei calculated using the unitary correlation operator method \cite{Ro06} with the Argonne V18 interaction. Upper part: Hartree-Fock approximation; lower part: with perturbative inclusion of long-range correlations.}
\label{fig:8}
\end{minipage}
\end{figure}

Of particular interest in this context is the observed systematics of nuclear properties through isotopic chains. This presumably reflects the isospin dependence of the underlying interactions governed by pions as chiral isovector Goldstone bosons. 
As a representative example, Fig.\ref{fig:7} shows calculated ground state deformations for several series of isotopes in comparison with experimental data. These results give some confidence that extrapolations into unknown territory of extreme nuclear isospins may be well guided by principles of chiral dynamics. Such extrapolations are required when addressing questions of astrophysical interest (nucleogenesis and r-process).  

Complementary approaches to finite nuclei, such as the unitary correlation operator method (UCOM)\cite{Ro06}, use different techniques but come to similar conclusions as to the importance of the tensor force\footnote{See also the reviews by T. Otsuka and D.J. Dean at this conference.}. The UCOM starts from a correlated many-body state $|\Psi\rangle = \exp[-iG]|\Phi\rangle$. Tensor correlations figure prominently in the correlation operator $G$. With long range correlations added perturbatively to a Hartree-Fock calculation, results such as the one shown in Fig.\ref{fig:8} are achieved. 

4.4~{\it Short-range correlations}.  Investigations of short-range $NN$ correlations in nuclei have a long history, with mixed conclusions drawn over past decades. Recent theoretical work has focused again on the high momentum components ($p > p_F$) in nuclear wave functions induced by strong tensor correlations \cite{SWPC07}. Major progress is now reported from dedicated experiments such as E01-015 at Jefferson Lab, designed to measure differences in correlations between proton-proton and proton-neutron pairs. Given the prominent role of the isovector $\pi + 2\pi$ exchange tensor interaction in the isospin $I = 0$ $NN$ channel, one expects strong short-distance correlations for $pn$ pairs und much weaker ones for $pp$ pairs. This effect is now clearly observed in the ratio of differential cross sections, $d\sigma(e,e'pn)/ d\sigma(e,e'pp) = 18\pm 4$ on $^{12}C$ measured at missing momenta between 0.3 and 0.6 GeV/c \cite{Sh07}.

4.5~{\it The third flavor dimension.} Adding strange quarks opens up a whole new chart of hypernuclear systems. The physics of $\Lambda$-hypernuclei is a well established branch of science. Nevertheless, a still persisting key issue is the weakness of the $\Lambda$-nuclear spin-orbit interaction \cite{Aj01} in comparison with the much stronger $L\cdot S$ force encountered in ordinary nuclei. Hypernuclear research will experience its next boost at the J-PARC facility.

Another fundamental question concerns the possible existence of quasibound antikaon-nuclear composites, given that the $\bar{K}N$ interaction close to threshold is known to be strongly attractive. However, the present data situation is unclear, and the predictive power of theoretical estimates is so far quite limited. The search goes on. This is another important case for the upcoming experimental program at J-PARC. 

\section{The nucleon: a many-body system full of surprises}

With its three valence quarks and an indefinite number of quark-antiquark pairs, all imbedded in strong gluon fields, the nucleon is the prototype system which exhibits all of the striking phenomena of low-energy QCD: notably confinement, spontaneous chiral symmetry breaking and its connection with the non-trivial structure of the vacuum. A key question concerns the origin of the nucleon mass: how do almost massless $u$ and $d$ quarks and massless gluons cooperate dynamically to form a localised baryonic compound with a mass of almost 1 GeV? An equally fundamental issue is the origin of the nucleon spin:  how is the total angular momentum of the nucleon in its rest frame distributed between its quarks and gluons, and in turn between their spin and orbital angular momentum?  

\begin{figure}[htb]
\begin{minipage}[t]{65mm}
\includegraphics[width=6.5cm]{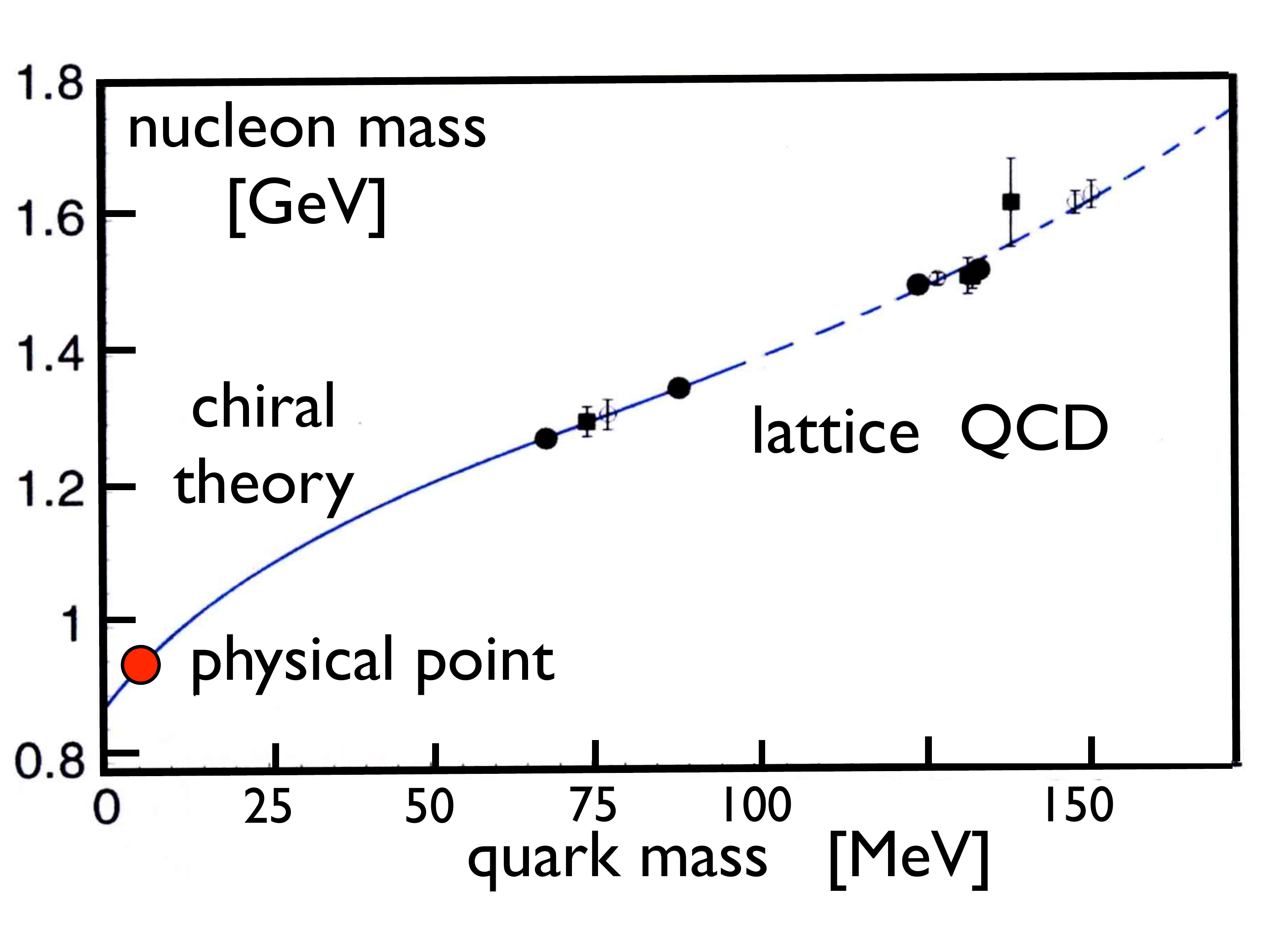}
\caption{Nucleon mass from lattice QCD and its interpolation to the physical point using chiral effective field theory \cite{PMW06}.}
\label{fig:9}
\end{minipage}
\hspace{\fill}
\begin{minipage}[t]{65mm}
\includegraphics[width=6.5cm]{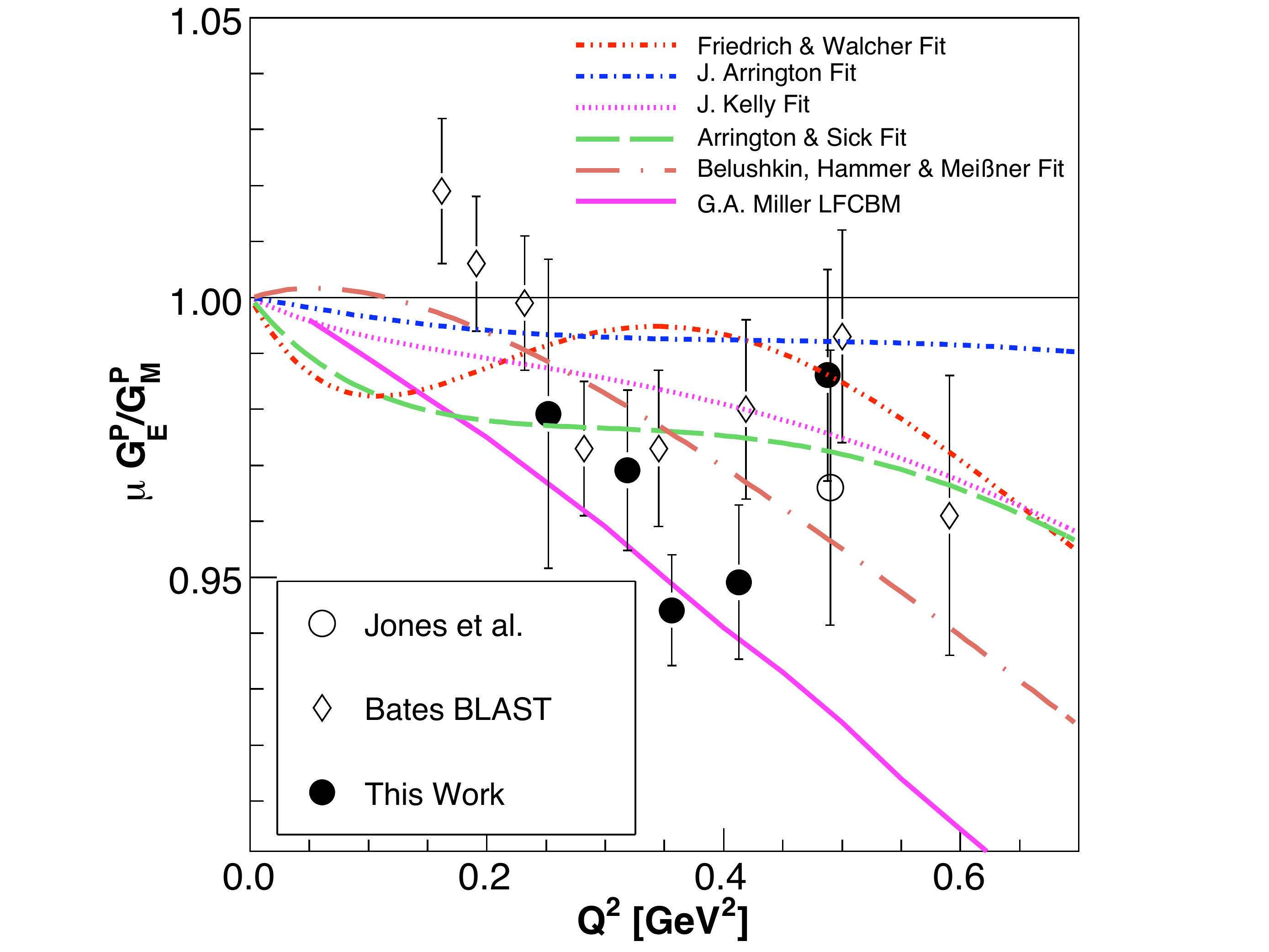}
\caption{Ratio of electric to magnetic proton form factors at low $Q^2$ measured with polarisation transfer $\vec{e} + p\rightarrow e + \vec{p}$ \cite{Ron07}.}
\label{fig:10}
\end{minipage}
\end{figure}

5.1~{\it The nucleon mass}.  Almost all of the mass of the visible universe is determined by the mass of the sum of the masses of nucleons in the cosmos. The gluonic energy density in the presence of three localised valence quarks obviously plays a decisive role in generating the nucleon mass.  Lattice QCD has progressed to the point that it can give reliable results concerning this issue, but with input quark masses still typically an order of magnitude larger than the actual current quark masses entering the QCD Lagrangian. Chiral effective field theory offers a systematic way to interpolate between lattice results and the physically relevant range of quark masses. 
Combining chiral perturbation theory with lattice QCD has thus become a widely used routine in recent years.  An example for the nucleon mass \cite{PMW06} is shown in Fig.\ref{fig:9}. Uncertainties are still sizable, given that the gap between the lowest equivalent pion masses (about 500 MeV) used in lattice simulations and the actual physical pion mass is still large. But as lattice QCD advances towards pion masses comparable to about twice the physical one, chiral perturbation theory begins to be a quantitatively reliable tool in conducting extrapolations down to the physical point.   

5.2~{\it Spin structure}. The present standards in disentangling the quark and gluon contributions to the nucleon spin, $1/2 = \Delta \Sigma/2 + L_q + \Delta g + L_g$, are set by the HERMES \cite{Ai07} and COMPASS \cite{Ag06} experiments. Only about one third of the nucleon's spin is carried by its quark constituents \cite{Ai07}, i.e. $\Delta\Sigma \simeq 0.3 - 0.4$. It appears now that the gluon contribution $\Delta g$ to the nucleon spin is small \cite{Ag06}. If this is confirmed with sufficiently high precision, it raises the interesting next question, namely about the contribution $L_q$ from quark orbital angular momenta. First lattice QCD results on moments of generalised parton distributions \cite{Hae07} indicate that the orbital pieces from $u$ and $d$ quarks come with opposite signs and roughly cancel to give $|L_q| \lesssim 0.04$. So the nucleon spin puzzle still persists.   

5.3~{\it Form factors}.  A further surprise came with the measurement of the ratio of proton electric and magnetic form factors, $\mu_P G_E(Q^2)/G_M(Q^2)$, in a polarisation transfer experiment ($\vec{e} + p\rightarrow e + \vec{p}$ ) at JLab \cite{Ga02}. This ratio, normalised to unity at $Q^2 = 0$, falls linearly with increasing $Q^2$, down to 0.5 at about 0.4 GeV$^2$, a behaviour not expected from the previous time-honoured Rosenbluth analysis. A good part of this apparent discrepancy is now understood in terms of two-photon exchange corrections which affect the Rosenbluth analysis much more than the polarisation transfer data. At very low $Q^2$ there may well be room for more surprises. In a re-analysis of the existing data base of electromagnetic nucleon form factors, a systematic irregularity was pointed out \cite{FW03} which had hitherto not been recognised. A dedicated low-$Q^2$ search performed at JLab \cite{Ron07} gives further hints of such fluctuations (see Fig.\ref{fig:10}) which are being discussed in terms of the pion cloud, the "soft" surface of the nucleon, but so far with no definitive conclusion.

\section{Hadrons in matter}

The (chiral) symmetry breaking scenario of QCD suggests that hadron masses vary with changing thermodynamic conditions. The mass gap seen in the spectrum of the lightest hadrons, which separates the non-trivial vacuum and its Nambu-Goldstone boson sector from all other hadronic excitations, is expected to depend on temperature $T$ and baryon density $\rho$. This presumably reflects the $T$ and $\rho$ dependence of the QCD condensates, and in particular, of the chiral condensate $\langle\bar{q}q\rangle$. Exploring changes of hadron properties, both in a ``normal" nuclear environment and in hadronic matter under more extreme conditions of temperature and density, is thus a persistingly relevant theme. 

Nucleons are well known to experience such changes in nuclei. Their effective {\it in-medium} mass is considerably lower than the one in free space. Their magnetic moments and weak decay matrix elements are significantly renormalised in a nuclear environment. The question is then whether mesonic excitations of the QCD vacuum and their spectra also show characteristic dependences on temperature and density. 

6.1~{\it Nambu-Goldstone bosons in matter}. By its Goldstone boson nature, the pion has its charge-averaged mass protected against in-medium changes. The pion decay constant $f_\pi$, on the other hand, varies approximately like the square root of the chiral condensate as it changes with increasing nuclear density. Detailed analysis of s-wave pion-nuclear interactions \cite{KKW03,FG07} in comparison with precision measurements of deeply bound states of pionic atoms \cite{Su04} and highly accurate data from low-energy $\pi^+$ and $\pi^-$ scattering on a series of nuclei \cite{Fr05} consistently display a systematics that can indeed be interpreted as showing the fingerprints of a decreasing chiral order parameter with increasing nuclear density.  

6.2~{\it Vector mesons in matter}. Vector mesons such as the $\rho$ and the $\omega$ are the lightest dipole-like quark-antiquark excitations of the QCD vacuum. Current algebra had established long ago a leading  proportionality between the $\rho$ mass in vacuum and the pion decay constant, $\sqrt{2}\,m_\rho = 4\pi f_\pi$, reflecting the relationship to the scale for spontaneous chiral symmetry breaking. The gliding of that scale with temperature on the way towards chiral restoration should then be observable through moments of the in-medium $\rho$ meson spectral function. This is the working hypothesis guided by a simple scaling argument (BR scaling \cite{BR91}) which has motivated this area of research ever since. 

\begin{figure}[htb]
\begin{minipage}[t]{65mm}
\includegraphics[width=6.5cm]{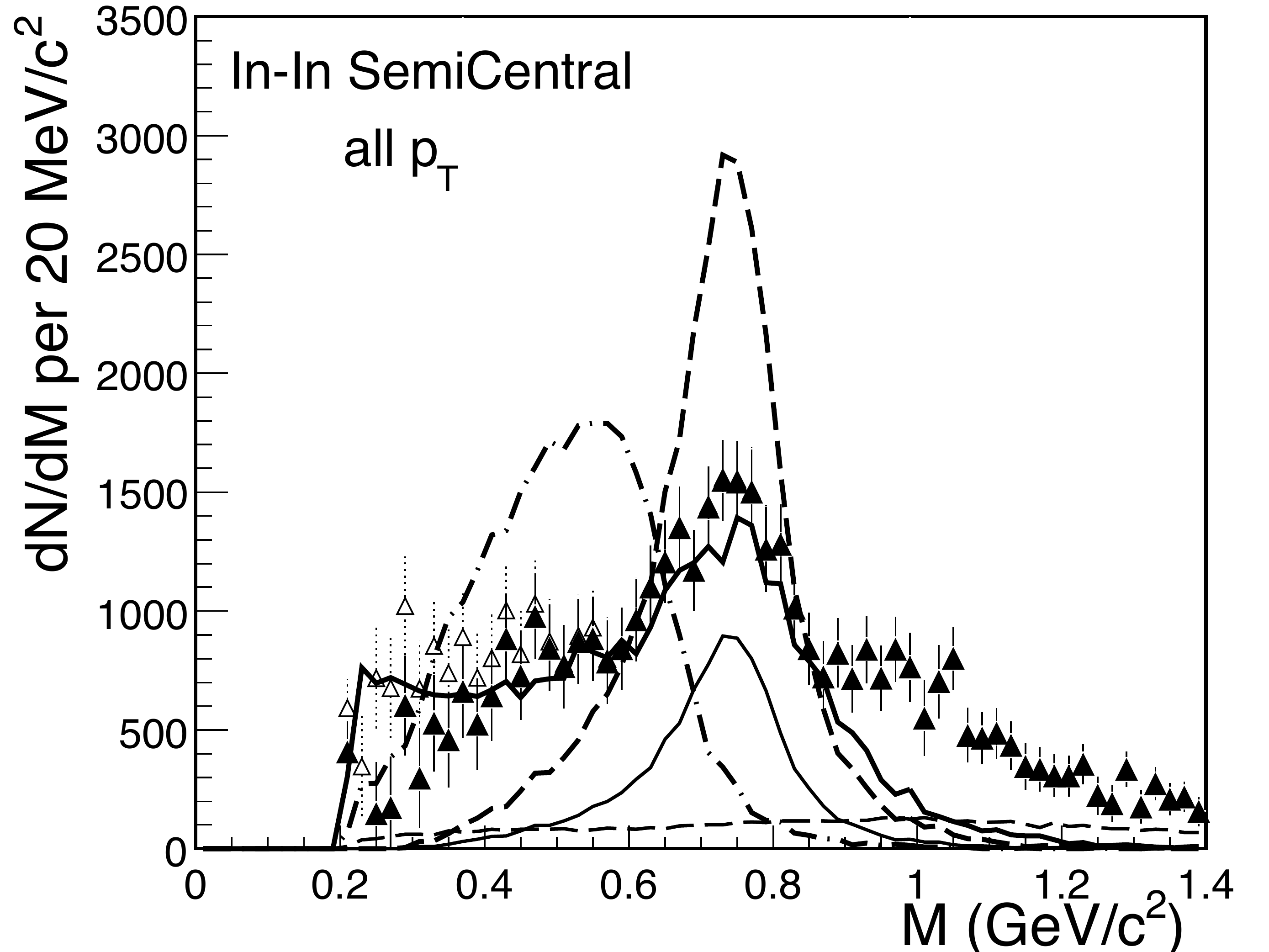}
\caption{Excess mass spectrum of muon pairs produced in 158 GeV In-In collisions at the CERN SPS \cite{Ar06}. The dashed curve shows the unmodified $\rho$ spectrum for reference. Other curves representing different calculations are explained in \cite{Ar06}.}
\label{fig:11}
\end{minipage}
\hspace{\fill}
\begin{minipage}[t]{65mm}
\includegraphics[width=6.5cm]{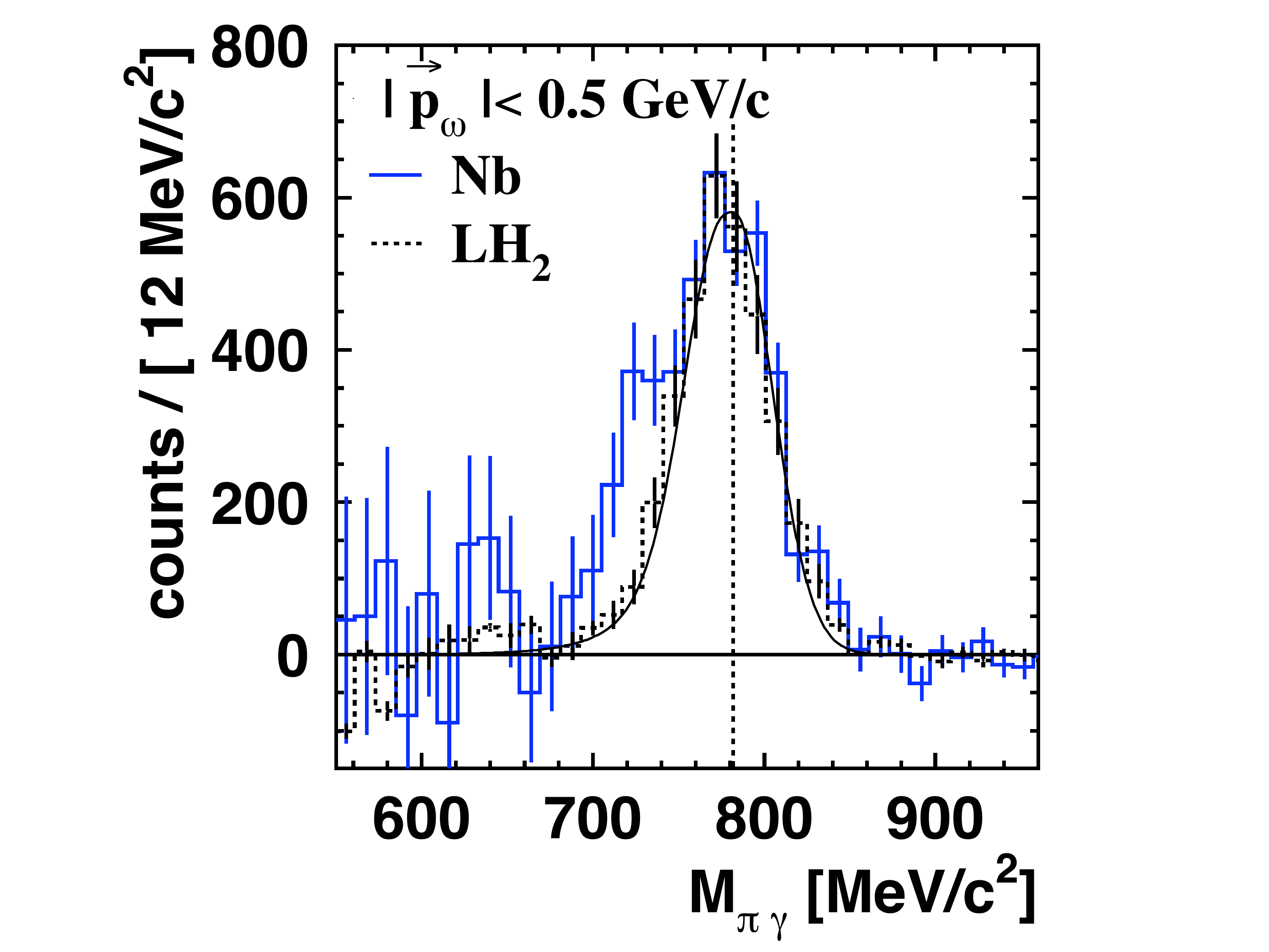}
\caption{Low-momentum $\pi^0\gamma$ invariant mass spectrum for photoproduction on Nb (solid) and LH$_2$ (dashed). The displacement from the free $\omega$ peak seen in the spectrum is interpreted as an in-medium mass shift consistent with $\delta m_\omega(\rho) \simeq -0.14\,m_\omega\,\rho/\rho_0$. (Taken from \cite{Tr05}).}
\label{fig:12}
\end{minipage}
\end{figure}

Where do we stand now? The NA60 experiment at CERN has made tremendous progress in sharpening observations of a low-mass enhancement of spectral strength in muon pair yields from high-energy nuclear collisions. The space-time averaged in-medium $\rho$ spectral distribution deduced from these measurements \cite{Ar06} (see Fig.\ref{fig:11}) is consistent with the theoretically expected strong broadening of that spectral function. However, the conclusion also drawn in \cite{Ar06}, namely that the $\rho$ meson does not show any noticeable shift in mass, should be taken with caution: once a spectral distribution experiences a significant broadening, the notion of {\it mass} can only be given a meaning in terms of the first moment of that distribution. A detailed analysis along these lines must still be performed before definite statements can be made.

A downward mass shift and broadening of the $\omega$ meson in the nuclear medium was theoretically predicted already a decade ago \cite{STT97,KKW97}. The investigation of $\omega$ meson photoproduction on nuclei using the CB/TAPS experiment at ELSA in Bonn \cite{Tr05} does indeed indicate a shift and broadening in the spectrum of low-momentum omega mesons (see Fig.\ref{fig:12})
detected via their decay, $\omega \rightarrow \pi^0\gamma$, although some details concerning background subtraction are still under discussion. As a further step ahead, the intriguing question about the possible existence of quasibound $\omega$-nuclear states is presently being explored. 

\section{Matter under extreme conditions}

We turn now to the more extreme sections of the QCD phase diagram described at the beginning of this report. Strongly interacting matter at very high temperatures is produced in heavy-ion collisions at the highest available energies, presently at RHIC and in the future at LHC. Cold matter at extreme densities exists in the interior of neutron starts. Both areas of research are rapidly developing, with exciting and unexpected results to be reported.

7.1~{\it Matter produced at RHIC}. The hot and dense matter produced and investigated by the RHIC experiments (PHENIX, STAR, PHOBOS, BRAHMS; see reviews and refs. in \cite{QM06}) has displayed a number of quite surprising properties\footnote{See also the review by W. Zajc at this conference.}. Three cornerstones of information drawn from these measurements are: i) transverse energy distributions of produced particles at mid-rapidity; ii) jet quenching; iii) eliptic flow of different hadrons and its successful description using hydrodynamics.
These results consistently lead to the conclusion that the hot matter produced has enormous initial energy densities of $10 - 20 $ GeV$\cdot$fm$^{-3}$, well above the energy density expected at the deconfinement transition.
It behaves like strongly coupled quark-gluon matter, opaque to particles carrying color. And it resembles a nearly perfect fluid with extremely low viscosity. All these features have been quite unexpected and differ substantially from the previously envisaged picture of a ``simple" quark-gluon plasma.

Further suprises include extremely fast equilibration, and the fact that for the chemical freeze-out, i.e. the formation of the hadronic species along with the expansion and cooling of the initially hot matter, a thermal (grand canonical) description of hadron yields works very well \cite{ABS06}. Such an analysis using the simplest hadron gas model suggests an ``empirical" freeze-out temperature of about $160$ MeV at zero chemical potential. Recent lattice QCD results \cite{Ch07} indicate a range of crossover temperatures $T_c$ which tend to be somewhat higher. The dispute about possible systematic uncertainties (finite volume corrections etc.) is still ongoing.

All these phenomena and their possible relationship to the QCD phase diagram now require a more detailed understanding. The experimental and theoretical focus is on correlations and transport properties in the produced quark-gluon medium: diffusion constants, conductivities, viscosities, susceptibilties. And we are looking forward to further breakthroughs once the heavy-ion program at LHC will be running.

7.2~{\it Neutron stars.} Matter under extreme conditions has its second frontier in the low-temperature, high-density domain which is realized in the core of neutron stars. Various models were proposed in order to extrapolate baryonic equations of state up to several times the density of normal nuclear matter, the range of densities relevant to neutron star interiors \cite{LP07}. Realistic many-body calculations of neutron star matter in terms of standard nucleonic and mesonic degrees of freedom, with the proton fraction determined by beta equilibrium, lead to relatively stiff equations of state. Phenomenological models with scenarios involving hyperons, pion or kaon condensates, or more exotic forms of quark matter, commonly generate equations of state that are substantially softer under gravitational pressure and therefore place rather low upper bounds on neutron star masses. 

The most accurate determinations of neutron star masses come from observations of double neutron star 
binaries, with average masses in a narrow range around the canonical $1.4\, M_\odot$ (solar mass units). It cannot not be excluded that the specific evolutionary circumstances leading to double neutron star systems favor such a narrow band of masses. Further significant developments are reported from observations of binaries with a neutron star accompanied by a white dwarf companion. Such systems show a wider spread of masses, though with individually larger error bars attached (see the compilation in \cite{LP07}). There is presently a lively discussion about the possible existence of neutron stars with masses as large as $2\, M_\odot$. This would eliminate a whole class of more exotic, soft equations of state. However, caution must be exercised in view of still existing uncertainties. Continued investigations, reducing the observational errors, are necessary to clarify the situation.  

\section{Concluding remarks}

We close this guided tour through various parts of the QCD phase diagram, the area of research that we identify as contemporary nuclear physics, with high expectations that many of the open issues mentioned in this brief report will be resolved or at least further clarified at the new and upcoming facilities (RIKEN, MAMI-C, JLab, J-PARC, FAIR, LHC, ...).



\end{document}